\documentclass[12pt]{iopart}
\usepackage[normalem]{ulem}  
\usepackage{graphicx}
\usepackage{capt-of}
\usepackage{amssymb}
\usepackage{lineno}

\def \FIGCOLORTYPE {_col}

\newcommand{\fid}{\texttt{FID}}
\newcommand{\lun}{\texttt{LUN}}
\newcommand{\fidstart}{\texttt{FID\_START}}
\newcommand{\lunstart}{\texttt{LUN\_START}}
\newcommand{\iic}{I$^2$C}

\newcommand{\ee}[2]{\ensuremath{#1\times10^{#2}}}
\newcommand{\pow}[2]{\ensuremath{#1^{#2}}}
\newcommand{\eg}{{\em e.g.}}

\begin{document}

\title{An absolute calibration system for millimeter-accuracy APOLLO measurements}

\author{E.G.~Adelberger$^1$, J.B.R.~Battat$^2$, K.J.~Birkmeier$^3$, N.R.~Colmenares$^4$, R.~Davis$^4$, C.D.~Hoyle$^5$, L.~Huang Ruixue$^2$, R.J.~McMillan$^6$, T.W.~Murphy,~Jr.$^4$, E.~Schlerman$^2$, C.~Skrobol$^3$, C.W.~Stubbs$^7$, A.~Zach$^3$}
\address{$^1$ Center for Experimental Nuclear Physics and Astrophysics, Box 354290, University of Washington, Seattle, WA 98195-4290, USA}
\address{$^2$ Department of Physics, Wellesley College, 106 Central St, Wellesley, MA 02481, USA}
\address{$^3$ TOPTICA Photonics AG, Lochhamer Schlag 19, 82166 Graefelfing / Munich, Germany}
\address{$^4$ Center for Astrophysics and Space Sciences, University of California, San Diego, 9500 Gilman Drive, La Jolla, CA 92093-0424, USA}
\address{$^5$ Department of Physics and Astronomy, Humboldt State University, One Harpst St, Arcata, CA 95521-8299, USA}
\address{$^6$ Apache Point Observatory, 2001 Apache Point Rd, Sunspot, NM 88349-0059, USA}
\address{$^7$ Department of Physics, Harvard University, 17 Oxford St, Cambridge, MA 02318, USA}
\eads{\mailto{tmurphy@physics.ucsd.edu}, \mailto{jbattat@wellesley.edu}}

\maketitle

\begin{abstract}
Lunar laser ranging provides a number of leading experimental tests of
gravitation---important in our quest to unify General Relativity and the
Standard Model of physics. The Apache Point Observatory Lunar Laser-ranging
Operation (APOLLO) has for years achieved median range precision at the
$\sim2$~mm level.  Yet residuals in model--measurement comparisons are an
order-of-magnitude larger, raising the question of whether the ranging data
are not nearly as \emph{accurate} as they are precise, or if the models are
incomplete or ill-conditioned.  This paper describes a new absolute
calibration system (ACS) intended both as a tool for exposing and
eliminating sources of systematic error, and also as a means to directly
calibrate ranging data in-situ.  The system consists of a
high-repetition-rate (80\,MHz) laser emitting short ($<10$~ps) pulses that
are locked to a cesium clock. In essence, the ACS delivers photons to the
APOLLO detector at exquisitely well-defined time intervals as a ``truth''
input against which APOLLO's timing performance may be judged and
corrected.  Preliminary analysis indicates no inaccuracies in APOLLO data
beyond the $\sim 3$~mm level, suggesting that historical APOLLO data are of
high quality and motivating continued work on model capabilities.  The ACS
provides the means to deliver APOLLO data both accurate and precise below
the 2~mm level.
\end{abstract}

\section{Introduction}

Lunar Laser Ranging (LLR) has long produced superlative tests of general
relativity (GR), using the solar system as a dynamical laboratory
against which the theory---or parameterized variants thereof---may be
tested \cite{Bender:1973zz,Dickey:1994zz,Williams:1976zz}.  The technique currently
provides our best tests of the strong equivalence principle (SEP),
time-rate-of-change of the gravitational constant, geodetic precession,
gravitomagnetism, and the inverse-square law, among others.  A recent
review article summarizes the science case for LLR
\cite{murphyLLRReview2013}.

For scale, post-Newtonian effects on the Earth-Moon separation, as
evaluated in the solar system barycenter (SSB) frame, appear at the 10~m
level.  Likewise, a complete violation of the SEP (\eg, if Earth's
gravitational self-energy did not itself experience gravitational
acceleration) would result in a 13~m amplitude range signal
\cite{Nordtvedt:1968zz,Damour:1995gi}.

Early LLR efforts achieved $\sim 200$\,mm measurement uncertainty, thereby
constituting tests of GR effects at the few-percent level.  The period from
1985--2005 saw an improvement in measurement uncertainty to the $\sim
20$~mm level, permitting $\sim 0.1$\% checks on relativistic gravity.
Beginning in 2006, the next level of LLR precision was achieved by the
Apache Point Observatory Lunar Laser-ranging Operation (APOLLO
\cite{murphyAPOLLO2008}), working down to the $\sim 2$~mm precision level
\cite{battatPASP2009, apolloCQG2012}.  Since coming online, APOLLO has produced the majority
of LLR measurements worldwide, and also habitually acquires ranges to 4--5
reflectors in each $\lesssim 1$~hr observing session.
In principle, this significant improvement in measurement precision should push
tests of relativistic gravity by a similar factor, down to the $\sim 0.01$\%
level in short order.  

However, LLR science relies on an intricate and
specialized model capable of reproducing all physical effects that can
influence the measurement from a telescope on Earth's surface to a
reflector placed on the lunar surface.  Few such models exist in the world
\cite{Fienga2015,Muller:2012sea,Williams:2012nc,Pitjeva:2013fja,Reasenberg:2017zjo}.
The basic scheme performs a numerical integration of the solar system from
a set of initial conditions (parameters in the model), following a
relativistic equation of motion that may also be parameterized to explore
deviations from GR.  Torques from non-spherical mass distributions are
included, and a host of ancillary effects are layered on top, such as
tidal distortion, Earth orientation, plate motion, and crustal loading.  The difference
between measured ranges and computed ranges form ``residuals,'' which are
iteratively minimized in a least-squares process by adjusting the unknown model
parameters (sometimes within known bounds).

If measurement uncertainties are properly assessed, and the model contains
all requisite physics (and is correctly coded), then the residuals should
distribute around zero according to the uncertainties such that a reduced
chi-squared measure would come out near unity.  Another way to express this
is that the weighted standard deviation (root-mean-square, or RMS) of the
residuals should be commensurate with the measurement uncertainty.

This is not the case for any of the current LLR models.  Residuals for
APOLLO's $\sim 2$~mm scale measurements are characterized by a spread
ranging from about 15~mm in the best model to about twice this in others \cite{apolloCQG2012,FiengaIWLR2016,HofmannIWLR2016,YagudinaIWLR2016}.
No obvious similar patterns emerge when comparing residuals from different
models.  Thus, introduction of LLR data an order-of-magnitude more precise
than was previously available has not yet significantly moved the needle in
testing gravity.

The situation raises questions such as: Are the data truly improved?  Are
the models capable of performing at the millimeter level?  Has the
technique run into fundamental modeling limitations?  Does the past data
collection effort of APOLLO represent ``money in the bank,'' establishing a
baseline of high-quality measurements while awaiting ultimately successful
model improvements---or does it indicate that resources were expended based
on imagined gains?

Efforts to improve models are ongoing, but years have passed without a
clear ruling on whether the large residuals indicate problems on the model
side or data side.  Evidence is mixed.  Performing several measurement
cycles around the lunar reflectors in the course of an hour shows a
consistency in APOLLO data points commensurate with estimated uncertainties \cite{battatPASP2009}
(while often showing separations by reflector as an indication that the
model was not getting lunar orientation right), but assessing possible
longer-term trends has proven difficult.  Meanwhile, measurements of a
second corner cube prism at the telescope exit aperture---in addition to
the ever-present fiducial corner cube prism---indicates measurement offsets
that depended on axial location, but at least in a seemingly static way
across months and years.  We suspect that the local corner cube (fiducial)
measurement may be impacted by the ring-down of the laser fire, which
involves switching $\sim 3000$~V in a matter of a few nanoseconds and thus
creating electromagnetic interference (EMI) that can impact the nearby
detector and timing electronics.

In light of these questions and concerns, we devised a calibration scheme
that allows us to independently assess APOLLO's data accuracy.  The
initial idea centered on injecting very short pulses onto the detector at
well-controlled intervals as a way to investigate APOLLO's timing
performance and track down sources of systematic error.  The concept
evolved into a high-repetition-rate fiber laser locked to a cesium clock,
whose pulses could be selected at will and delivered to the APOLLO system
by a long fiber, thus isolating the system from the APOLLO-generated EMI.

Most powerfully, we can overlay these calibration photons atop the lunar
range measurements, acting as a calibrated ``optical ruler'' delivering photon
tick marks to the detector alongside lunar return photons.  Thus,
we are able to calibrate the measurements directly, even if we never manage
to find and eliminate sources of systematic error.  In so doing, we
transform APOLLO from a few-millimeter \emph{precision} apparatus to a
few-millimeter \emph{accurate} operation.

This paper describes the Absolute Calibration System (ACS), and what we
have learned thus far about APOLLO data quality.  The conclusion is that we
see unappreciated influences at the $\sim 3$~mm level, but nothing that
would account for the $\gtrsim 15$~mm scale of model residuals.  The
implication is the most beneficial to the scientific community: past APOLLO
data are of high quality, so that as soon as model capabilities catch up,
this long baseline of few-millimeter-accurate data can be leveraged for
improved constraints on gravitational physics.

\section{ACS Overview and Design}

The original concept for the ACS involved a pulse-on-demand laser system
capable of delivering photons to the APOLLO detector at prescribed times,
in pulses that would have a high degree of relative accuracy (in the time
interval between pulses). Such a setup would permit a comparison of
APOLLO-determined time intervals to the ``truth'' data from the ACS pulse
pairs. Any offset could be studied in a variety of conditions, including
various delays after the main APOLLO laser fires. Equipment tunings, module
swaps, shielding, temperature effects, etc. could all be quantified in
terms of their impacts on APOLLO accuracy.

We had difficulty identifying a pulse-on-demand laser operating at
or near the APOLLO wavelength of 532~nm having a pulse width narrower
than about 90~ps. Moreover, commercial delay generators that could produce pairs
of triggers several seconds apart (simulating the lunar round trip
travel time) could not provide better than 50~ps jitter.
Finally, these devices were not easily configurable
to multiplex interleaved pulse pairs in order to faithfully mimic
LLR measurements at high rates. While some of these barriers may have
yielded to solutions, the combination of challenges motivated us to explore
alternate ideas, culminating in the more powerful ACS design described here.

Fiber-cavity lasers employing a saturable absorber mirror naturally produce
picosecond pulses at high repetition rates. We adopted a laser design
that generates 1064~nm $\sim 10$\,ps pulses at 80\,MHz---and most
importantly, is capable of modulating the cavity length in a phase-locked
loop referenced to a frequency standard such that the period could be
stabilized. Intrinsic stabilization can result in pulse jitter well below
one picosecond. For reference, 1\,mm of one-way distance translates to
6.7\,ps of round-trip time. When paired with a cesium standard, the
interval of time $\Delta t$ between pulses could be made extremely
regular. This scheme does not support placement of a pulse at
arbitrary time, but the tradeoff is that a pair of pulses separated by
several seconds will have a reliable $\Delta t$ limited only by the
clock jitter. For a high-performance cesium standard, this jitter is
typically better than 10\,ps: roughly an order of magnitude better
than the pulse-on-demand timing uncertainty. As a result, we rely much
less on statistical averaging to get below the millimeter
limit, so that in principle the requisite statistics may be accumulated in a few
seconds, and the system timing stability can be readily investigated.

This scheme also lends itself well to the second, potentially far more
powerful application of the ACS: we can inject calibration photons
during lunar ranging operations, knowing that all pulses are tightly
connected to a high-performance clock. Thus we deliver an optical
clock to the avalanche photodiode (APD) array detector, overlaying a ``ruler'' of ``tick marks''
simultaneous with the lunar range measurement. If the fiducial corner
cube measurement is being skewed by electromagnetic noise following
the APOLLO laser fire, for instance, the ACS calibration will tell us
by how much, effectively in real-time. In this way, it would matter less if we
never successfully delivered on the first function of the ACS: finding
and eliminating sources of systematic error. Naturally, it is
important that we make every attempt to do so, but we can still
realize a transition from precision to accuracy (in a more guaranteed
manner, in fact) by employing this overlay technique---now standard practice
in APOLLO data acquisition.

\subsection{ACS implementation}
The ACS has three main sub-systems:  (a) the ``ACS Enclosure'' containing the 80\,MHz 1064\,nm pulsed laser, the Cs clock, and associated control systems, (b) a pulse processing system that selects and conditions specific pulses from the train and transmits them to (c) the APD delivery optics.
The ACS Enclosure is located away from the main APOLLO system, and is
electrically isolated from the rest of the ACS and APOLLO apparatus. The
ACS laser pulse train is delivered to the pulse processing system via 12~m
of polarization-maintaining single-mode optical fiber.  The ACS Enclosure
does not move in elevation with the telescope, unlike the APOLLO laser and
electronics and the remainder of the ACS system.  A pair of Raspberry Pi
computers is used to manage the interface between the ACS and the rest of
the APOLLO apparatus.

\subsubsection{ACS laser and clock overview}
\begin{figure}[t]
  \centering
  \includegraphics[width=0.95\textwidth]{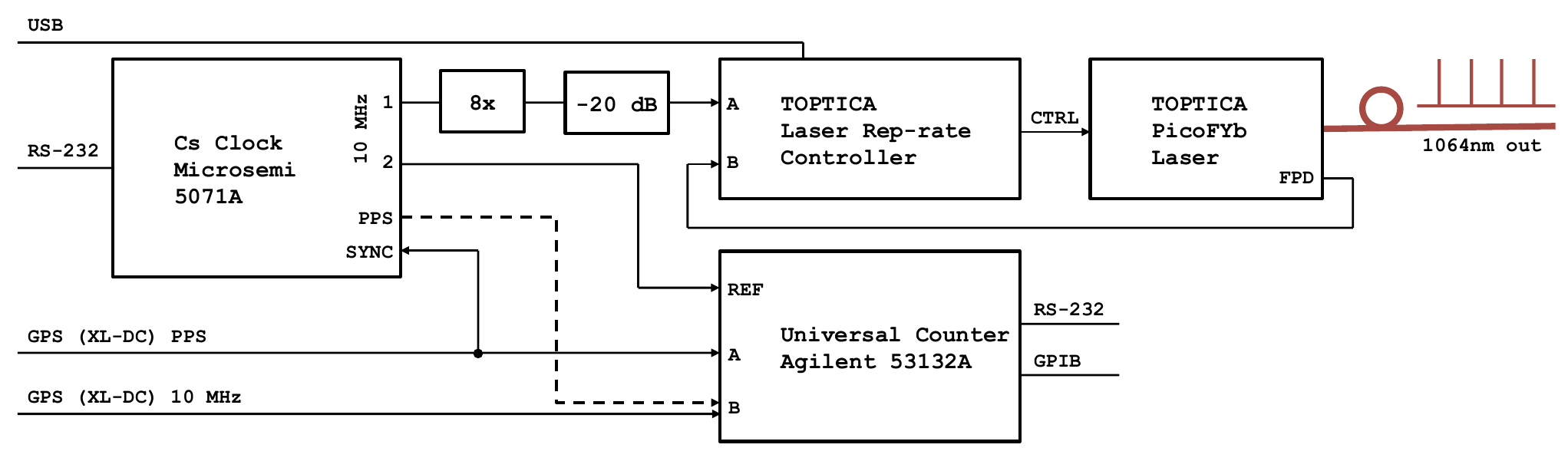}
  \caption{\label{fig:acs_enc} The ACS Enclosure contains a laser (TOPTICA
PicoFYb) that delivers a series of 1064\,nm 10~ps pulses into an
optical fiber. A Laser Repetition-rate Controller (LRC) locks the fiber
laser to a Cs frequency standard (Microsemi 5071A). A Universal Counter
(UC: Agilent 53132A) enables a comparison between the Cs clock and a
separate GPS-disciplined clock (TrueTime XL-DC).
The dashed line represents occasional and brief reconfigurations to monitor
the accumulating phase difference between the two clocks, as discussed in
the text.}
\end{figure}

The ACS Enclosure (Figure~\ref{fig:acs_enc}) houses the Cs frequency
standard (Microsemi 5071A; stability in Table~\ref{tab:allandev}), which
provides a pair of equivalent 10\,MHz 1\,V RMS sinusoidal signals having
low phase noise ($< -155$~dBc at high frequency). Phase jitter at the
important 2.5~s lunar round-trip time computes to about 2.5~ps.  One of the
10~MHz outputs from the Cs clock provides the 80\,MHz reference for the
ACS laser, after being frequency multiplied by a factor of 8 (Wenzel
IFM-3R-10-8-13-13) and attenuated to match the input expectation of the
loop controller. The frequency multiplication increases the phase noise by
$20\log 8\approx 18$~dB, and the phase jitter at 2.5~s periods becomes
about 6.3~ps.

\begin{table}[tbh]
  \caption{\label{tab:allandev}Allan deviation of the 5071A Cs clock, as provided by the manufacturer.}
  \begin{indented}
  \lineup
  \item[]\begin{tabular}{@{}llllll}
    \br
    Delay [s]       & \pow{10}{0} & \pow{10}{1} & \pow{10}{2} & \pow{10}{3} & \pow{10}{4} \\
    Allan Dev. & \ee{3.1}{-12} &  \ee{2.1}{-12} & \ee{5.9}{-13} & \ee{2.0}{-13} & \ee{6.9}{-14} \\
    \br
  \end{tabular}
  \end{indented}
\end{table}

The ACS laser is a turn-key system that can be operated remotely.  It is a
customized TOPTICA PicoFYb that is diode-pumped and uses an all-fiber
(Ytterbium-doped) configuration based on a ring oscillator with a
semiconductor saturable absorption mirror (SESAM).  The laser pulse width
at the output of the 12\,m optical fiber is $\sim 10$\,ps (full width at
half-maximum: FWHM), delivering 30~mW average power within a wavelength
envelope of 0.5~nm. Further details about the ACS laser are provided in
Section~\ref{sec:acs_laser}.

The relative frequency and phase difference between the Cs and
GPS-disciplined (XL-DC) clocks are monitored by a Universal Counter (UC,
Agilent 53132A).  The Cs 10\,MHz signal serves as the frequency reference
for the UC, and the frequency of the XL-DC clock is measured relative to
this reference to 1~$\mu$Hz resolution in $\sim 10$~s gates. In an
alternative configuration (requiring a cable exchange, see dashed line in
Figure~\ref{fig:acs_enc}), the time difference between the Pulse-Per-Second
(PPS) signals from the two clocks is measured to a resolution of 0.1~ns.
The results of this comparison are summarized in
Section~\ref{sec:clock_comparison}.

\subsubsection{Pulse processing \label{sec:pulses}}

\begin{figure}[t]
  \centering
  \includegraphics[width=0.9\textwidth]{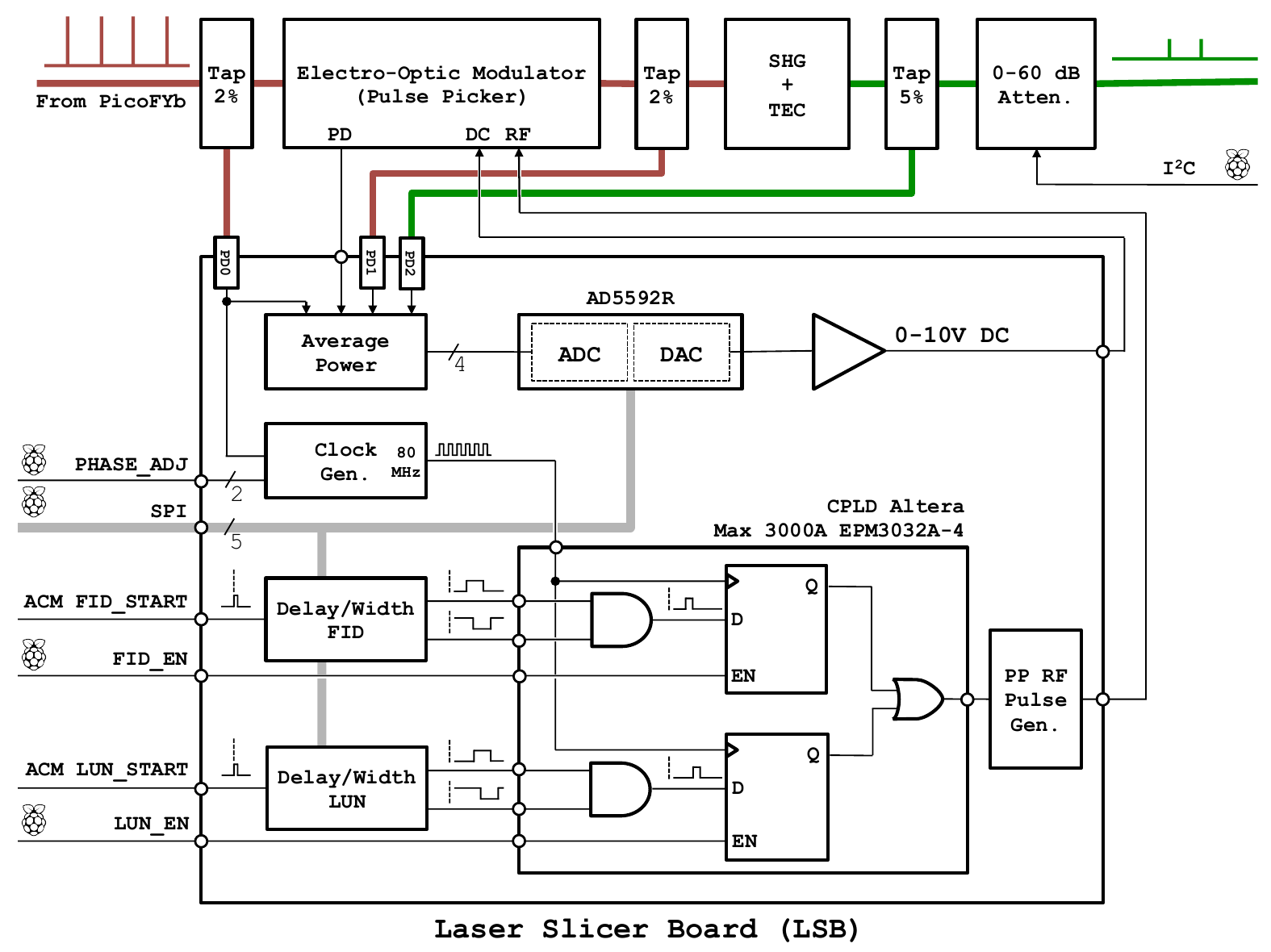}
  \caption{\label{fig:lsb} Schematic diagram of the pulse processing system
that selects, conditions, and delivers specific ACS laser pulses to the APD
injection optics. Thick red and green lines represent optical fibers, while
the thin black lines indicate electrical connections. The ACS pulse train
is modulated by an Electro-Optic Modulator (EOM), which is normally held in
the blocked (``off'') state.  The Laser Slicer Board (LSB) generates RF gate pulses
to make the EOM transmissive for one or more pulses.  Details about the LSB
and gate pulse generation are provided in the text. A Second Harmonic
Generator (SHG, thermally regulated by a Thermo-Electric Controller, TEC)
frequency-doubles the light to green (532\,nm).  A variable attenuator
(0--60\,dB) allows remote control of the pulse amplitude, generally tuned
to deliver $\sim 1$ ACS photon per pulse to the APD array.  Three optical
taps enable the LSB to monitor the average laser power along the signal
path, and also to generate an 80~MHz clock slaved to the pulse train.
Signals labeled with \includegraphics[height=10pt]{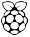} indicate
connections to a Raspberry Pi (GPIO or \iic{} or SPI).}
\end{figure}

Injecting calibration photons at critical times requires the ability to
select individual pulses out of the 80~MHz pulse train.  
We accomplished this with a custom pulse processing system (see
Figure~\ref{fig:lsb}).  The ``pulse slicing'' is performed by an
Electro-Optic Modulator (EOM), or ``pulse picker,''  which blocks most pulses
and passes only the desired ones onward to a temperature-controlled Second
Harmonic Generator (SHG).  In the SHG, a non-linear optical element
converts pairs of 1064\,nm photons into single 532\,nm photons.  Next, the
green laser pulses pass through a remote-controlled variable attenuator set
to deliver $\sim 1$ photon per pulse to the APD array. The EOM is
controlled by the Laser Slicer Board (LSB), which monitors gate request
signals from the APOLLO Command Module (ACM, not shown), and generates
EOM-compatible windowing pulses with user-defined delays and widths.

Using on-board photodiodes, the LSB monitors the
average laser power before and after the EOM (1064\,nm: PD0 and PD1,
respectively), and after the SHG (532\,nm: PD2).  The EOM contains an
internal photodiode whose average photocurrent is also measured by the LSB.
Potentiometers tune the sensitivity of each channel to effectively utilize
the full range of the 12-bit analog-to-digital converter (ADC: AD5592R).

The LSB generates an 80~MHz clock signal to facilitate synchronized
pulse-picking (described below) based on the PicoFYb laser
pulses sensed by PD0 (see \texttt{Clock Gen.} in Figure~\ref{fig:lsb}).
The clock phase, relative to the optical pulses, is controlled by a PLL-based
clock multiplier/divider with a programmable delay line (Texas Instruments
CDCF5801A), described in greater detail below.

\begin{figure}[t]
  \centering
  \includegraphics[width=0.75\textwidth]{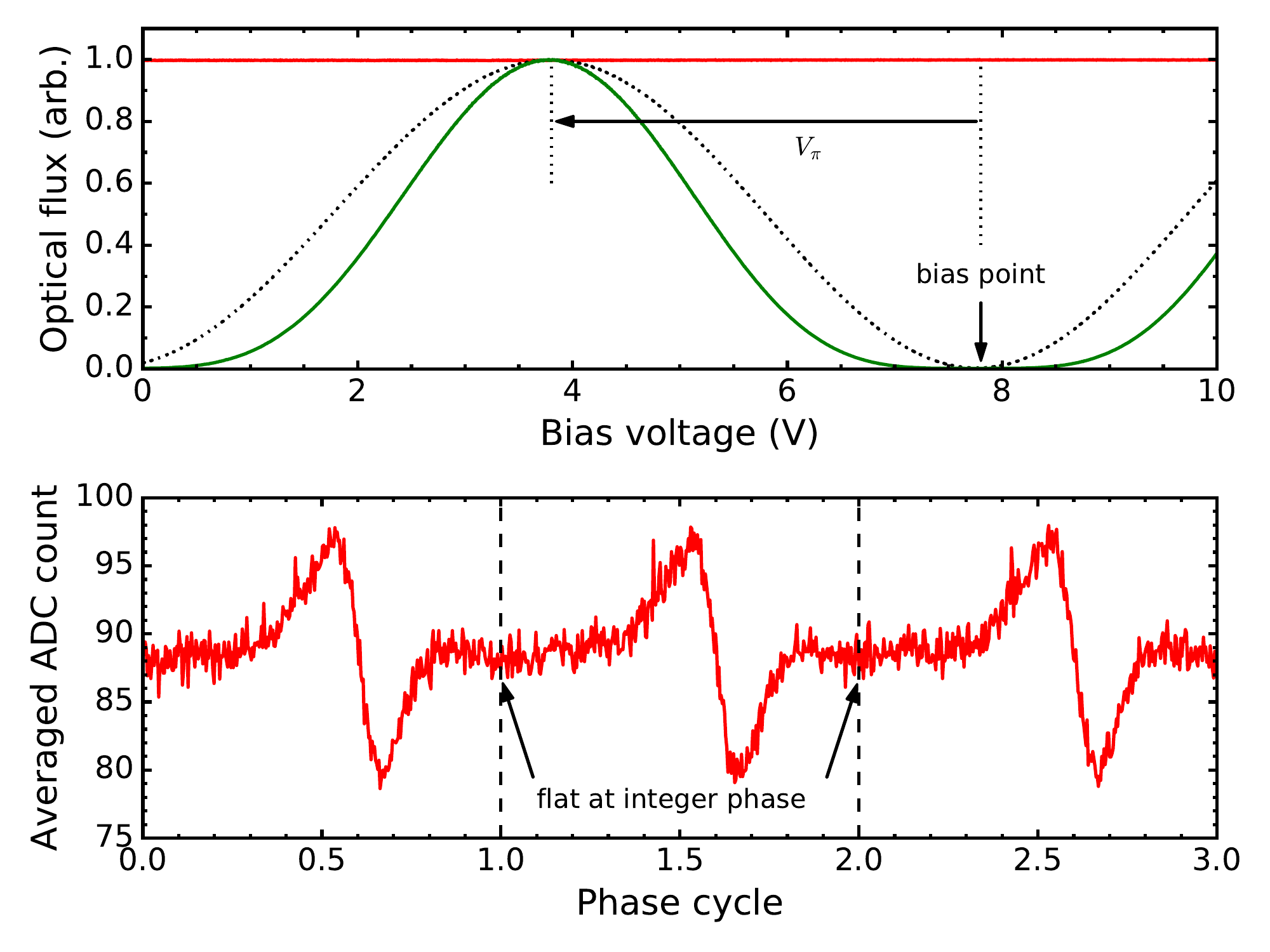}
  \caption{\label{fig:sweeps} Routine sweeps establish the optimal DC bias of the EOM (top) and phasing so that selection edges are far from optical pulses at the EOM (bottom). At top is a sweep of the DAC output controlling
the pulse-picker DC bias. The flat, solid (red) curve at top is the IR input
(PD0, steady; unaffected by pulse picking); the black dotted curve is the IR
pulse-picker output (PD1); solid (green) is the post-SHG output (PD2), essentially the
square of its input.  We typically bias the EOM near the 8~V minimum and switch by amplitude $V_\pi$.  At bottom is the PD1 signal during a sweep through
three cycles of the LSB-generated 80\,MHz clock phase.  The hump and trough
features arise from having the selection window edges coincident with
optical pulses. To select pulses cleanly, we adjust the phase so that
integer values find flat regions between the edge features.}
\end{figure}

The EOM (Photline NIR-MX-LN-10-PD-P-P) is a fast (10~GHz) Mach-Zehnder
interferometer in which the index of refraction of one arm, and therefore
the relative phase of the light in the two interferometer arms, can be
controlled by an external voltage. A change in bias voltage of $V_\pi \sim
4$\,V will toggle the EOM from destructive to constructive interference,
with an extinction ratio of 30\,dB (on top of the $\sim 5$\,dB insertion
loss).  The quadratic behavior of the SHG results in $\sim 60$~dB switching
contrast in green light.  Although the EOM could be used as a variable
attenuator, we operate it as a two-state device: maximally opaque
(``blocked'') or maximally transmissive (``open''). The LSB generates a DC
bias for the EOM via a DAC (Digital-to-Analog Converter in the AD5592R,
followed by an amplifier), tuned to hold the EOM in the blocked state. The
blocked-state DC bias voltage is measured experimentally by sweeping the DC
bias voltage from 0 to 10\,V and monitoring the average laser power after
the EOM (PD1).  Figure~\ref{fig:sweeps} (top) shows an example sweep.  The
value of $V_\pi$, the voltage difference between the blocked and open
states, can be verified from a DAC sweep.  The DAC sweep also allows a
measurement of the EOM extinction ratio ($\sim 10^3$) by comparing the
average laser power in the blocked and open states (the ratio of the
maximum and minimum readings of PD1).

The following paragraphs describe the scheme for clean selection of
individual laser pulses.  The overall idea starts with a logic signal
initiating the request.  Programmable delay chips define a user-controlled
period during which 80~MHz clock pulses generated from the ACS laser pulse
train may trigger a synchronous pulse request.  To avoid the transmission
of runt or double optical pulses, the gate pulses must bracket the optical
laser pulses to ensure that the EOM is opened before the desired laser pulse
arrives, and returns to the blocked state well in advance of the next
pulse.  The amplitude of this RF pulse is set to $V_\pi$ (manual
potentiometer on the LSB) in order to flip the EOM state from blocked to
open for one or more pulses.

In more detail, the EOM gates that pick individual pulses are generated in
response to a trigger from the ACM, which is part of the APOLLO timing
system.  Each firing of the main APOLLO laser generates two gated detector
events: one for the local (fiducial) corner cube return and another for the
eventual lunar return.  We refer to these as \fid{} and \lun{} gates,
utilizing logic trigger pulses at the beginning of each gate type called
\fidstart{} and \lunstart{}, as labeled in Figure~\ref{fig:lsb}.
Considering a single gate type for what follows, a pulse-request window is
created by first stretching the trigger pulse to $\sim 120$~ns, then
passing to a pair of programmable delay chips (Maxim DS1023-50; 128~ns
range in 0.5~ns steps), generating two differently delayed copies of the
input pulse.  An inverted version of the more-delayed signal is
AND-combined with the earlier, non-inverted version to create a composite
pulse whose beginning and end times are user-controlled at the 0.5~ns level
for precise definition of the EOM window. This combined pulse serves as the
input to a D-type flip-flop, so that if the logic signal is high when a
positive-going clock edge arrives, a 12.5~ns pulse request is exported. In
this way, the gate request edges are guaranteed to be multiples of 12.5~ns
apart, at a constant and controllable phase relative to the laser pulse
train.  The AND and flip-flop are implemented in a complex programable
logic device (CPLD, Altera Max 3000A EPM3032A-4), also employing an
OR-combination of both \fid{} and \lun{} pulse requests so that separate
timing may be defined for each.  Considering propagation delays, we are
able to generate pulse requests as early as 27~ns after the initial trigger
edge (\fidstart{} or \lunstart{}) emerges from the ACM and as late as
155~ns, in 0.5~ns steps.

Correct phasing of the EOM requests relative to the optical pulses is facilitated by the CDCF5801A chip, which allows bi-directional control of the phase 
using two GPIO lines from the Raspberry Pi (see \texttt{PHASE\_ADJ} in Fig.~\ref{fig:lsb}).
The direction of the phase adjustment is determined by the high/low state of one GPIO line,
while a phase step is triggered by a rising edge on the second GPIO line.
After 1536 rising edges, the clock phase cycles by 360$^\circ$ (0.234$^\circ$ of phase per step). 
We get confirmation that the EOM-enable gate is properly phased relative to
the laser pulse train by sweeping the phase and measuring the average
transmitted laser power (PD1) while selecting pulses at a roughly 1~MHz
rate. We identify and operate at a flat region in the output, away from
fluctuations indicating runt or double pulses when the phase is such that
the gate selection edges coincide with laser pulses at the EOM.  A sample
phase is sweep shown in Figure~\ref{fig:sweeps} (bottom).

\subsubsection{APD delivery optics}

\begin{figure}[t]
  \centering
  \includegraphics[width=0.9\textwidth]{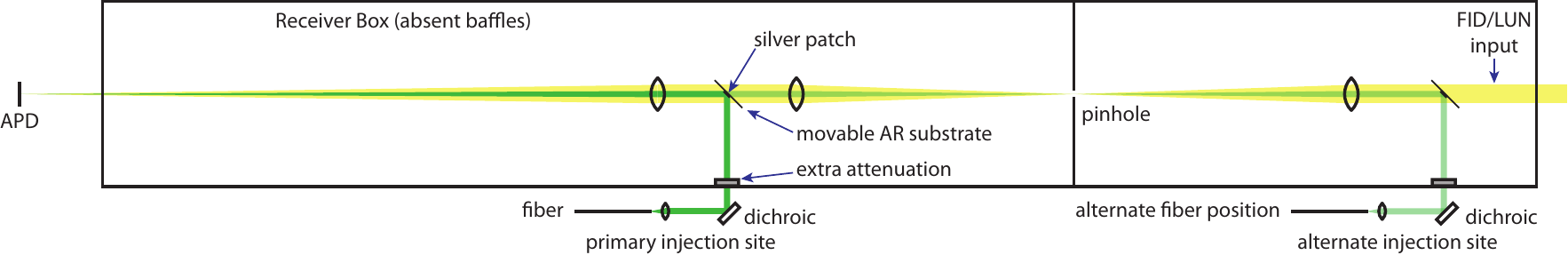}
  \caption{\label{fig:optics} Optical delivery of ACS photons to the APD
detector does not block lunar or fiducial photons from the telescope, as
ACS photons are situated near the APOLLO receiver optical axis in the
shadow of the telescope secondary mirror.  Additional attenuation
(presently $10^{-3}$) at the receiver box entry, in tandem with controlled
attenuation (Figure~\ref{fig:lsb}) establishes single-photon illumination
of the APD array. The alternate injection site is discussed in the text.}

\end{figure}

After pulse slicing, frequency doubling to 532\,nm, and intensity
adjustment, the ACS laser pulses must be delivered to the APD array without
blocking the lunar signal photons. Figure~\ref{fig:optics} shows the
optical system that does this. First, the light exits the fiber and a lens
creates a free-space collimated beam.  Next, a 45$^\circ$ dichroic mirror
reflects the 532\,nm light into the well-baffled APOLLO receiver box,
eliminating $> 96$\% of any residual 1064\,nm light that makes it through
the 532\,nm single mode fiber.  A neutral-density filter at the receiver
wall contributes to the overall attenuation of the system.  The ACS photons
are injected into one of two possible collimated sections in the receiver
box.  The central portion of this collimated section is effectively obstructed
by the secondary mirror of the telescope, meaning that a small reflective
patch placed in the center can couple the ACS photons into the APD
without blocking any APOLLO signal photons.  This is achieved by a
45$^\circ$ clear optic (anti-reflection coated for 532\,nm), with a small
silvered patch in the center.

The alternate injection site---seldom used---allows study of any impact to
system timing imposed by additional optics and attenuators (not all shown)
within the receiver.  In normal operation, rotating attenuator disks near
the pinhole impose a $\sim 10^5$ attenuation for \fid{} returns compared to
\lun{} returns, which would produce unequal ACS amplitudes for \fid{} and
\lun{} gates.

\subsection{ACS Laser}
\label{sec:acs_laser}

\begin{figure}[t]
  \centering
  \includegraphics[width=0.7\textwidth]{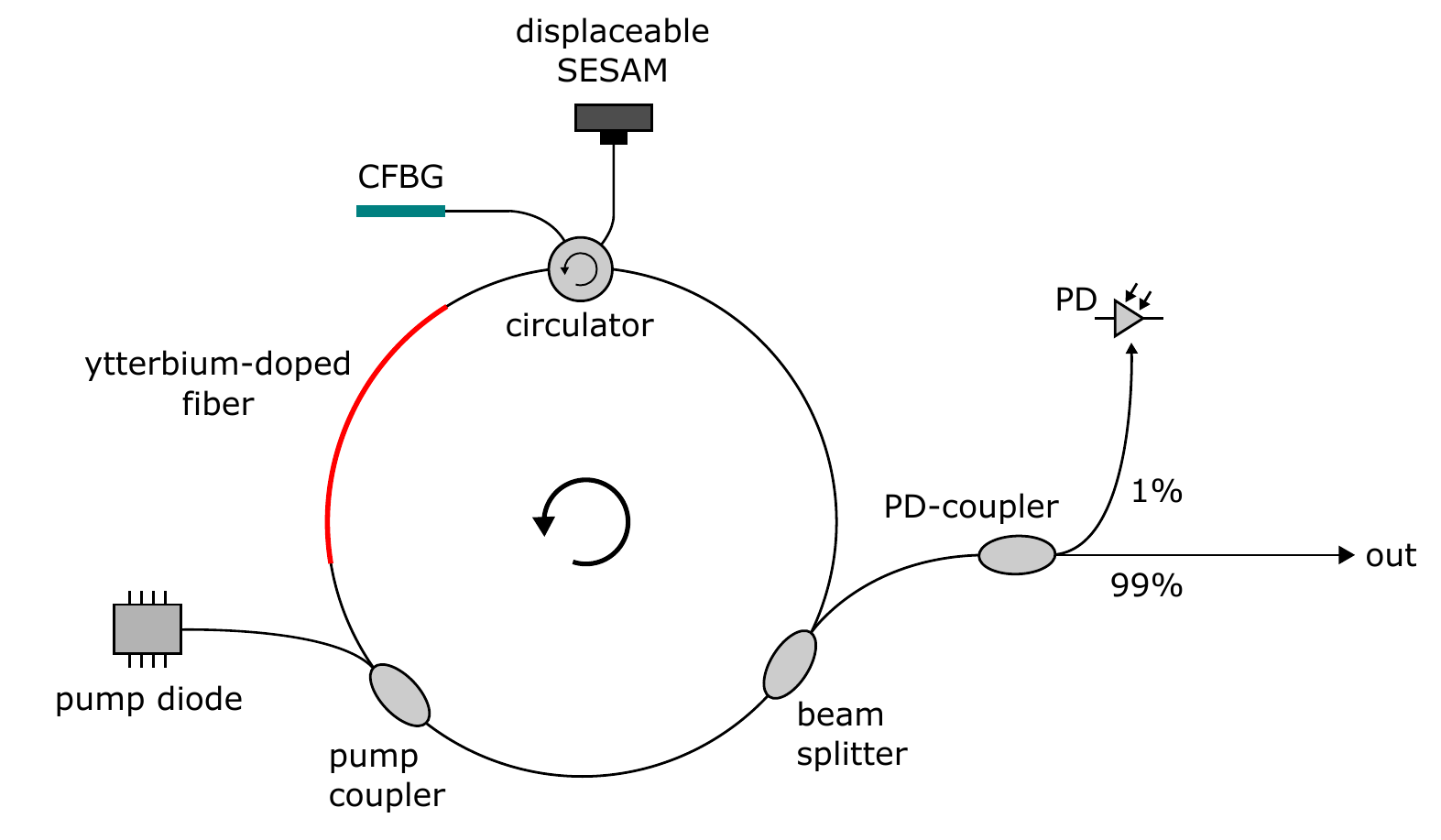}
  \caption{\label{fig:laser} Schematic diagram of the ring-shaped fiber
oscillator.  See text for details, and Ref.~\cite{laserPatent} for patent information.}
\end{figure}

The setup of the TOPTICA PicoFYb laser is depicted in Figure~\ref{fig:laser}. A
SESAM mode-locked fiber oscillator
is the starting point. The repetition rate is controlled and stabilized by
adjusting the cavity length with a piezo actuator. The complete fiber
oscillator is temperature controlled by Peltier elements.

The laser emits pulses at a central wavelength of $\lambda=1064.29$\,nm,
$\Delta\lambda\approx 0.191$\,nm (FWHM) and a pulse width of $\sim 10$\,ps---as measured via an autocorrelator using a sech$^2$ fit.  An average
power of about 35\,mW is coupled out of the oscillator at a repetition rate
of 80\,MHz.

About 1\% of the output power of the oscillator is split and detected by a
photodiode in order to generate the electrical signal for the phase-locked
loop (PLL) electronic control of the repetition rate (within the LRC in Figure~\ref{fig:acs_enc}). The instantaneous phase
difference between the Cs clock reference signal and the signal of the
photodiode is measured by an analog phase detector. Its output voltage
is proportional to the phase difference and is fed to a
proportional-integral-derivative (PID) regulator as an
error signal. The PID regulator minimizes the error signal by applying an
electrical voltage to the piezo actuator and hence varying the optical
length of the oscillator by displacing the SESAM along the beam axis. A
software-controlled regulation of the oscillator temperature compensates
for low-frequency drifts of the repetition rate. The phase detector
additionally delivers an RMS phase jitter signal that is used to determine
the residual RMS timing jitter value. 

A residual RMS timing jitter of 350\,fs is measured in closed-loop action
when integrated from 10\,Hz to 100\,kHz. This comparatively high jitter value
is considered to be mainly caused by the type of the laser oscillator which
has high anomalous net intracavity dispersion in order to operate in the
soliton regime and is therefore strongly affected by Gordon-Haus timing
jitter. Nevertheless, this oscillator type was chosen for stability reasons
and the achieved RMS jitter value is far below the minimum requirement for the
ACS of $< 2$\,ps. 

The pulse duration of 10\,ps---longer than a more natural $\sim 2$\,ps for
this type of laser---was chosen in order to minimize nonlinear effects in
the 12\,m PM980 fiber delivery following the laser.  In particular,
self-phase modulation (SPM) could have a detrimental effect on the
efficiency of the second harmonic generation (SHG) stage that follows.
After the fiber delivery, the pulse exhibits a slight temporal lengthening
of about one picosecond to $\sim 11$\,ps, and a spectral broadening by a
factor of 2 to $\sim 0.4$\,nm.

\begin{figure}[t]
  \centering
  \includegraphics[width=0.7\textwidth]{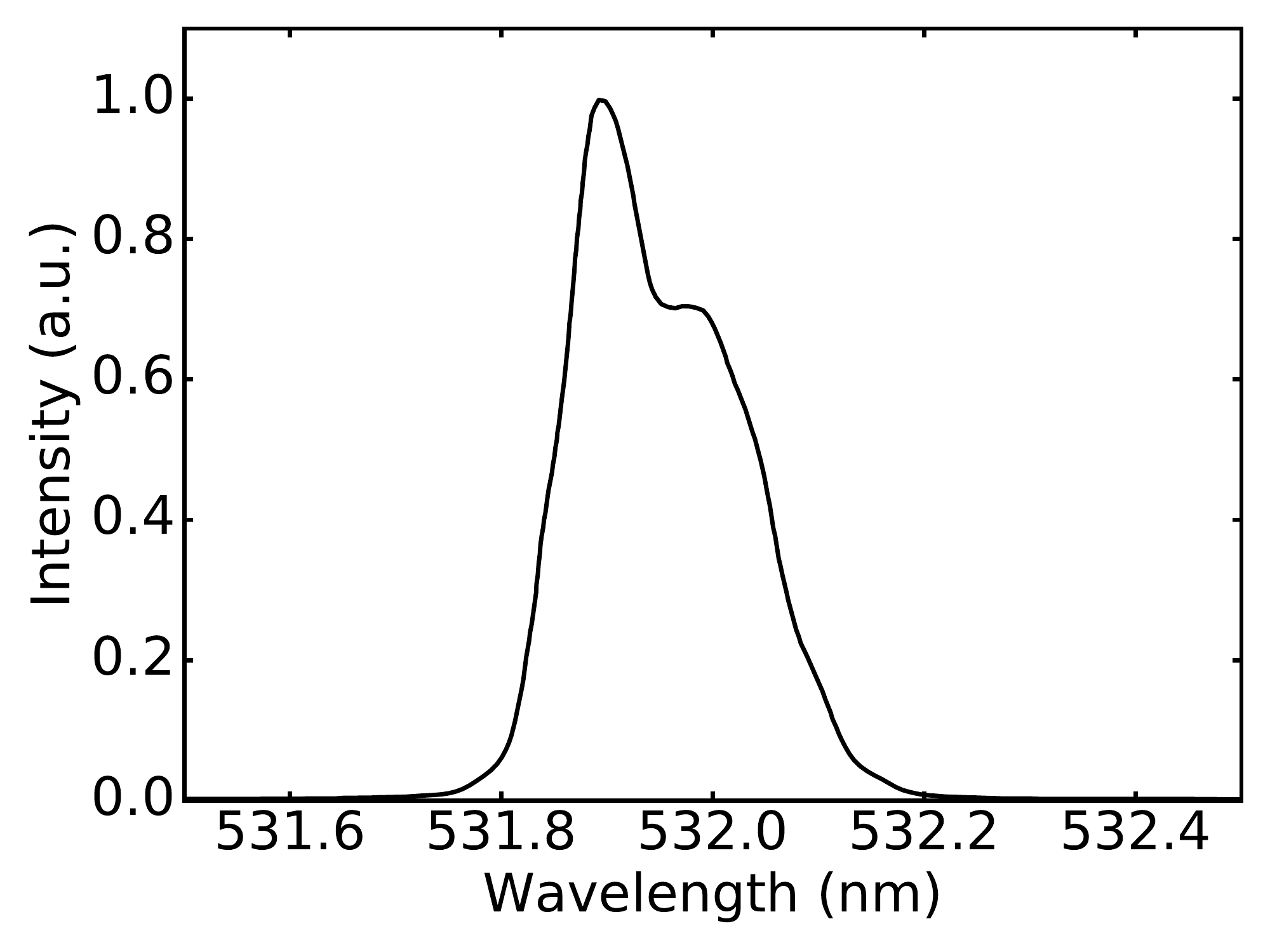}
  \caption{\label{fig:spectrum} Spectrum of the frequency-doubled pulses
at the output of the SHG.
Self-phase modulation in the 12\,m fiber delivery causes a slight drop at the central
wavelength.}
\end{figure}

The pulse picker introduces an optical loss of about 5\,dB.  Therefore a
fundamental power of $\sim 11$\,mW is used to seed the SHG unit. A
temperature-stabilized 5\,mm thick MgO:PPLN crystal is used for frequency
doubling to $\lambda\approx 532$\,nm (the spectrum is depicted in
Figure~\ref{fig:spectrum}). The frequency-doubled pulses contain an energy
of about 4\,pJ.  The low peak intensity of the fundamental pulse results in
a rather low SHG efficiency of 3\%, which is more than sufficient to
deliver the desired one calibration photon per pulse to the APD array.  It
was not possible to measure the temporal pulse profile because of low pulse
energy. Nevertheless, in theory the quadratic power dependency of the
nonlinear process in the non-saturated case causes a reduction of the pulse
length after the SHG to a pulse duration of $< 11$\,ps.  A dichroic filter
at the exit of the SHG aperture suppresses infrared transmission, letting
only 0.02\% through.  Roughly half of the green light is coupled into the
output PM480 fiber.

As implemented in the ACS system, we find approximate laser power and pulse
energy levels as detailed in Table~\ref{tab:laser-energy}.  The three taps
shown in Figure~\ref{fig:lsb}
deliver approximately 520, 220, and 9\,$\mu$W, respectively, if the EOM is
held in its most transmissive (open) state.  Individual green pulses
emerging from the fiber contain approximately 6 million photons.  The
variable attenuator is typically set to $-16$\,dB, and a fixed 3\,ND filter
is placed at the entrance to the optical receiver enclosure.  Expanding the
beam to uniformly cover the APD array adds another factor of ten of
effective attenuation.  Together with other losses and detection
efficiencies, we end up capturing approximately one photon per pulse across
the 16 elements of the APD array.  The contrast between green and infrared
photons at the APD array is always high ($>10^5$ in the open state and
$>10^2$ in the closed state) due to the combination of dichroic elements
and fiber coupling/transmission.

\begin{table}[t]
  \caption{\label{tab:laser-energy}Laser average power and pulse energy at various stages.}
  \begin{indented}
  \lineup
  \item[]\begin{tabular}{@{}lll}
    \br
    Stage & Power (mW)  & Energy (pJ)\\
    \mr
    Out of 12\,m fiber   & 30      & 375 \\
    After upstream tap   & 26      & 325 \\
    After (open) EOM     & 13      & 160 \\
    After downstream tap & 11      & 140 \\
    SHG fiber output     & \00.18  & \0\02.2 \\
    \br
  \end{tabular}
  \end{indented}
\end{table}

\section{ACS Results}

\subsection{Clock comparison}
\label{sec:clock_comparison}

A GPS-disciplined clock (TrueTime XL-DC model) previously served as the
sole time base for APOLLO. The Cs clock recently added as part
of the ACS provides superior frequency stability on intermediate timescales and can be used to characterize the GPS clock. Although the XL-DC clock
is no longer the frequency standard for APOLLO ranging, it continues to
operate as a time standard, referenced to Universal Time Coordinated
(UTC).  A detailed comparison of the two clocks, along with a method for
back-correcting a decade of APOLLO data based on recorded XL-DC clock
statistics is described elsewhere \cite{APOLLO_clock2017}.

On a roughly weekly schedule, we reconfigure cables into the UC for
approximately one minute in order to check the accumulated phase drift
between the two clocks.  At the 2.8\,km altitude of the site, we expect the
free-running Cs clock to run fast by $3.0\times 10^{-13}$ due to
gravitational redshift.  Our Cs clock was measured by the manufacturer
to have a static frequency offset of $-1.3\times 10^{-13}$ relative to a
hydrogen maser standard, for a net frequency offset from the UTC second of
$1.7\times 10^{-13}$.  The resulting phase accumulation should be roughly 15\,ns
per day.  Indeed, we measure a $\sim 14$\,ns per day drift between the two
clocks.  The net frequency offset translates to a lunar ranging measurement
error of less than 0.1\,mm.

\subsection{Example ACS run and products \label{sec:example-run}}

This section illustrates the gains delivered by the ACS, while providing
details on the techniques by which the results are realized. The next
section provides a first-look into the longer-term behavior of the ACS
calibration measures.

We use for demonstration a 10,000-shot (500 second) observation of the
Apollo~15 reflector on 2016-09-12---the first ever using the ACS.  While
this run is not fully representative in some ways (\eg, still using XL-DC
clock as frequency standard rather than Cs clock), it is perhaps more
instructive in that it illustrates the effects of clock drift and GPS
disciplining.  During this run, the fiducial gates and lunar gates each saw
about 20,000 total photon events (for clarity, a ``gate'' is a
detector activation event associated with laser ``shots''---a prompt one
for the local corner cube and another delayed to catch the lunar return).
Of these, we get roughly 11,000 useful ACS photons for each gate type,
while about 5,000 and 3,000 photons contribute to the fiducial and lunar return
peaks, respectively.  Raw timing data are in the form of 12-bit integer counts
(bins) from a time-to-digital converter (TDC) having 16
channels---representing pixels in the $4\times4$ APD array---and 25\,ps
resolution over 100\,ns. We will focus on the lunar gate in what follows,
but the fiducial gate is analogous.

\begin{figure}[t]
  \centering
  \includegraphics[width=0.5\textwidth]{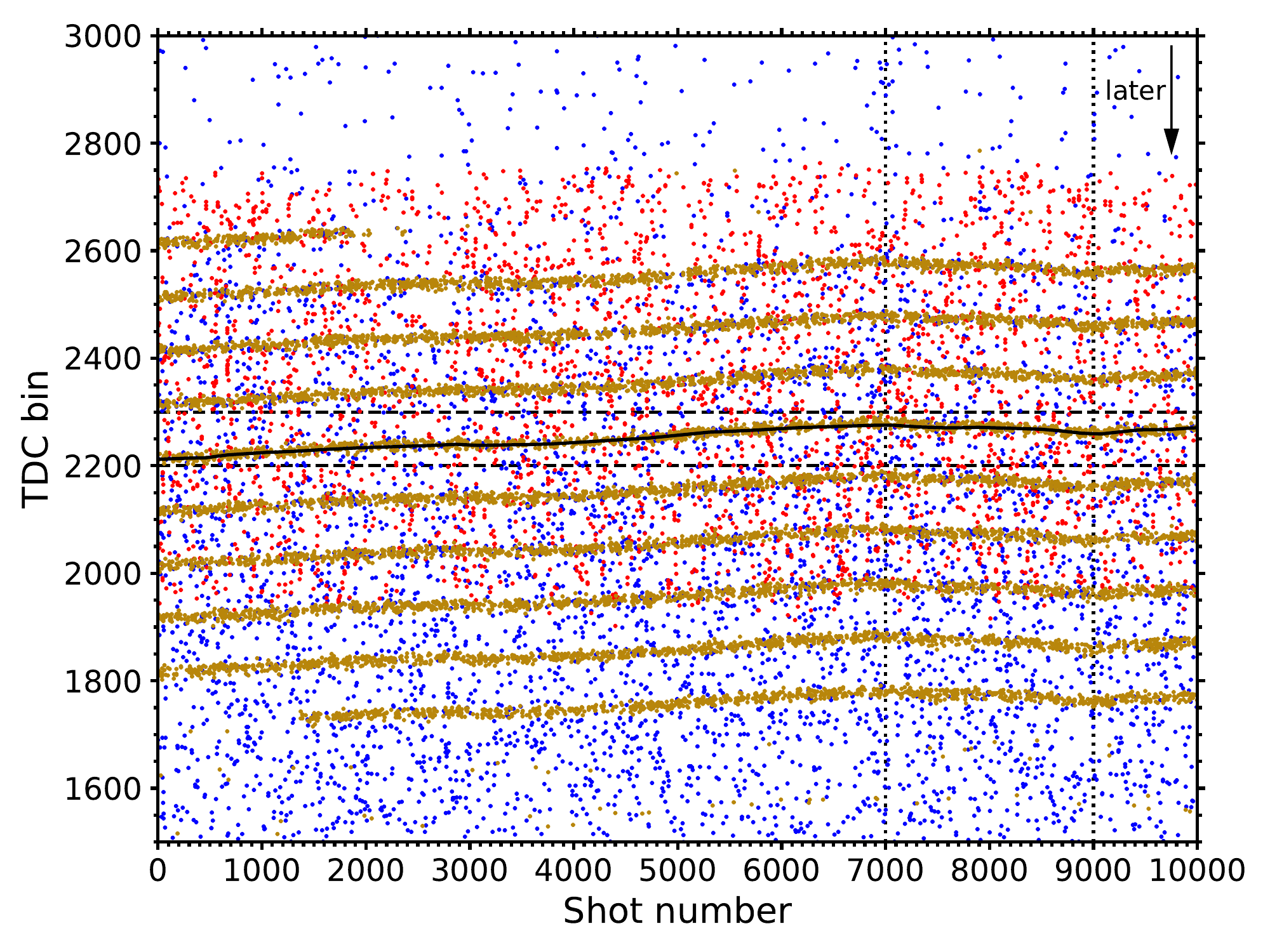}\hfill{}\includegraphics[width=0.5\textwidth]{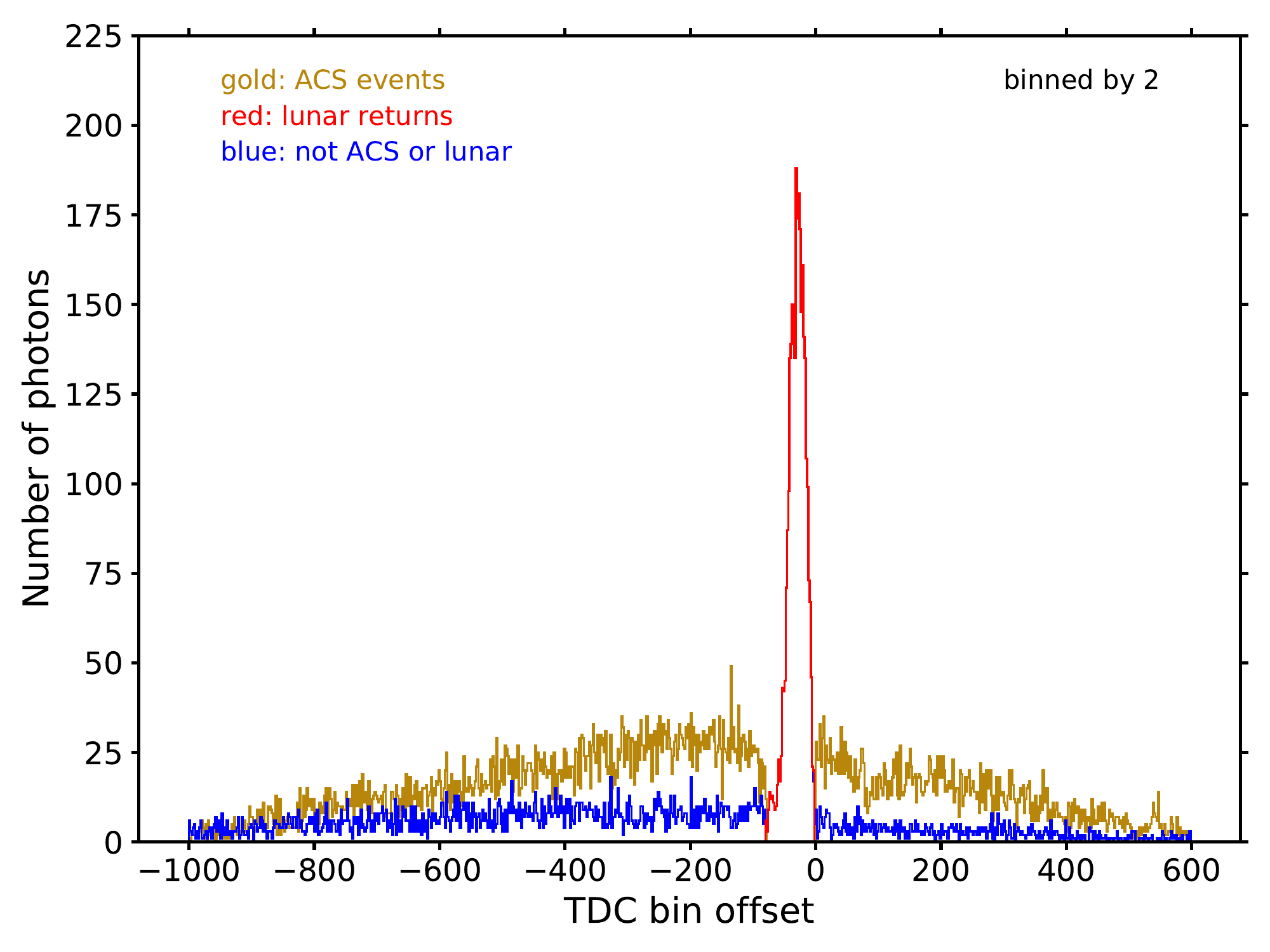}
  \caption{Left: raw data from the first LLR+ACS run obtained over a 500\,s
period on 2016-09-12. Each dot represents a photon detection during the
lunar gate. TDC time measurements are 25\,ps per bin; later photons appear
lower in the plot. Yellow dots have been identified as ACS pulses based on
nearly-static phase relative to the XL-DC clock. Red dots coincide with
predicted lunar return times. Blue dots are the remainder---ones that
cannot be confidently identified as either lunar or ACS---and largely
represent background, slow avalanches due to carrier diffusion, or delayed
crosstalk events in the APD (becoming more pronounced later/lower in the
figure). The solid black line is constructed from the independent UC
measurement of the APOLLO clock frequency referenced to the Cs clock.
Horizontal dashed lines help to emphasize the degree to which the two
clocks drift, and vertical dotted lines delimit the shot range during which
the XL-DC clock steering reversed direction for 100\,s. Right: histogram of
the lunar-prediction-corrected TDC values, smearing the ACS spikes into a
triangular distribution while pulling lunar returns into a high-visibility
signal---even though weaker than the aggregate ACS signal and spread over
approximately the same TDC region.  Color assignments match those in the
left panel.  The masking by ACS photons in the left panel looks five times
worse here than it really is, because only one in five stripes is ``live,''
or possible, for a given shot. \label{fig:lun_raw}}

\end{figure}

Raw TDC data (corrected only for individual static channel offsets
to improve visibility/alignment) appear in the left panel of Figure~\ref{fig:lun_raw},
while the right panel shows the same data, but rearranged by subtracting
the \emph{predicted} TDC value that a lunar return would have \emph{for
the corresponding shot}---thus piling up lunar detections and smearing
out ACS pulses. This is important to understand: based on laser fire
time, we can predict to better than a nanosecond (40 TDC bins) when returning lunar
photons will arrive. Because the return photons are asynchronous with
respect to the 50\,MHz APOLLO clock, they distribute uniformly over
a 20\,ns range in the TDC, which measures the time between the return
photon (START) and a selected subsequent clock pulse (STOP). But we know \emph{when},
in TDC-space, any lunar photon should appear. Those arriving at the expected
time are colored red in Figure~\ref{fig:lun_raw}.

\begin{figure}[t]
  \includegraphics[width=0.9\textwidth]{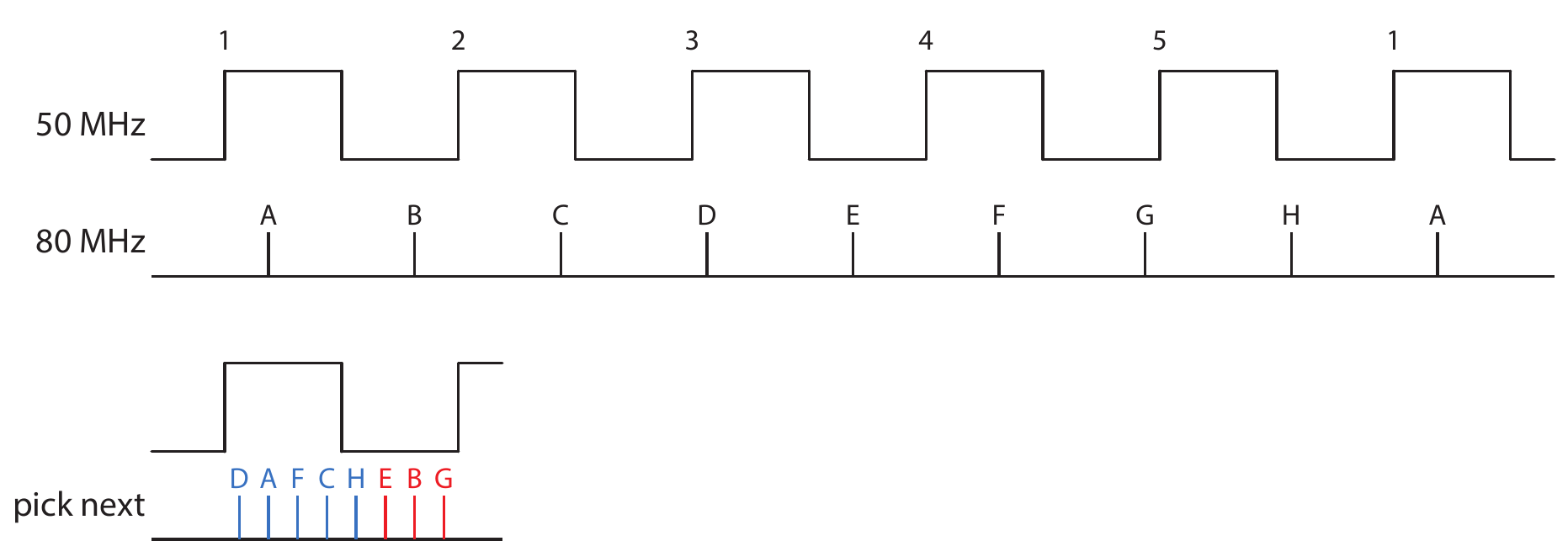}
  \caption{\label{fig:acs_phasing} Relative timing between 50\,MHz XL-DC clock and 80\,MHz ACS laser pulses. Absent frequency shifts between the signals, there are 5 possible alignments of the 80\,MHz laser pulse train relative to any given 50\,MHz rising edge (then the pattern repeats). The resulting comb of laser spikes has a 2.5\,ns spacing.}
\end{figure}

Note that the ACS photons (yellow) in Figure~\ref{fig:lun_raw} form
a ``comb'' of stripes 2.5\,ns (100 TDC bins) apart in the left panel,
overlapping the (red) lunar photons reasonably well (the overlap is
not perfect in this \emph{first} on-sky run, but we have since adapted
using fine control of ACS pulse positioning---as detailed in
Section~\ref{sec:pulses}). While ACS
pulses are actually integer multiples of 12.5\,ns apart, the 80~MHz pulse
train has five possible positions
relative to the 50\,MHz clock pulse that forms the TDC STOP measure (see
Figure~\ref{fig:acs_phasing}).  The number of stripes appearing in
Figure~\ref{fig:lun_raw} depends on the adjustable width of the pulse
selection window as described in Section~\ref{sec:pulses}.
The drift rate of the XL-DC clock (GPS-disciplined) relative to the
ACS clock (Cs) eventually spreads ACS photons across all TDC values.
This drift disappears when the Cs clock is used as the APOLLO
time base, as both the ACS laser pulses and TDC STOP pulses become
synchronous.

We can play many games using identified ACS photons to learn more
about APOLLO timing. The games we explore here \emph{all start like
this}: pick some shot: \eg, shot 1000. Consider each/any ACS photon
identified for that shot in the fiducial gate, for which we record
APD channel/pixel and TDC value. Now ask if there are any lunar-gate
ACS photons in the \emph{same} channel for the \emph{same} shot number
(or offset shot in some studies), roughly 2.5\,s later, when the lunar
return would arrive. Finally, ask: what time difference would APOLLO
report between these two events, using our standard techniques for
turning raw TDC measurements into time differences? How does this
compare to an integer multiple of 12.5000\,ns, assuming the ACS photons
represent absolute truth in timing? Now repeat this timing error
assessment for multiple shots, and possibly multiple channels, aggregating
results into a single distribution for statistical analysis. If better
statistics are wanted, one may also include shot numbers that are
not exact matches, but are within $\pm\delta_{\mathrm{shot}}$ of
the target (so that $N_{\mathrm{shot}}=2\delta_{\mathrm{shot}}+1$
lunar gates are considered for each fiducial shot/gate).

The more channels we include, and the wider the shot window ($\delta_{\mathrm{shot}}$),
the more photons we have for comparison. For example, using all 10,000
shots in the example LLR run, APD channel 16 exhibits 74 shots having
ACS photons in the same shot for both gate types. Allowing $\delta_{\mathrm{shot}}=5$
shots (\eg, lunar gates for shots 995 to 1005 compared to the fiducial
gate for shot 1000) results in 730 pairs to compare, and much better
statistics. Opening up to all channels provides 826 pairs when exact
shot/gate matches are required, and 8918 when $\delta_{\mathrm{shot}}=5$.
Some results populate Table~\ref{tab:example-stats}. Note that timing
errors are converted to range errors in millimeters in the round-trip
sense, wherein 1\,ps corresponds to 0.15\,mm. Timing and range errors
may be used interchangeably, in this sense.

\begin{table}[t]
  \caption{\label{tab:example-stats}APOLLO range errors are small, and can be determined precisely depending on how many channels and shots are considered.}
  \begin{indented}
  \lineup
  \item[]\begin{tabular}{@{}llll}
    \br
    Channel & $\delta_{\mathrm{shot}}$ & $N_{\mathrm{pairs}}$ & Range error (mm)\\
    \mr
    16    & 0 & \0\0\074 & $-1.70\pm3.58$\\
    16    & 5 & \0\0730  & $-0.13\pm1.09$\\
    1--16 & 0 & \0\0826  & $-0.55\pm1.01$\\
    1--16 & 5 & \08918   & $\m0.30\pm0.32$\\
    1--16 & 9 & 15234    & $\m0.47\pm0.24$\\
    \br
  \end{tabular}
  \end{indented}
\end{table}

\begin{figure}[t]
  \centering
  \includegraphics[width=0.6\textwidth]{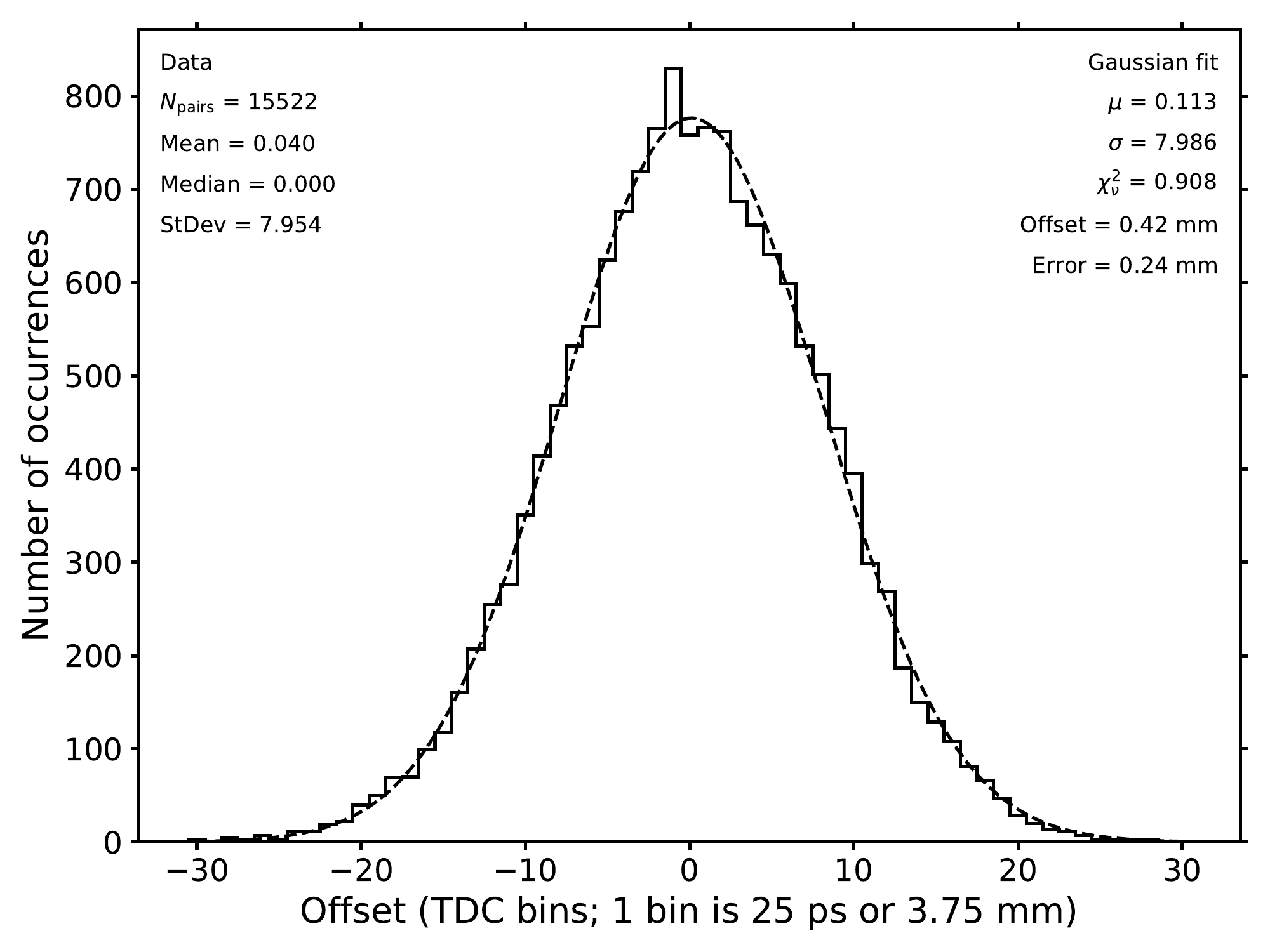}
  \caption{Example histogram corresponding to the final row in Table~\ref{tab:example-stats}.  The result is well-described by a Gaussian distribution with mean $\mu=0.126$ TDC bins (0.47\,mm of one-way path), and standard deviation $\sigma=7.94$ TDC bins.  The error, computed from $\sigma/\sqrt{N_{\rm{pairs}}}$, is 0.24\,mm of one-way path.\label{fig:gaussian}}
\end{figure}

The principal product of the tables in this section is the range error in
the final column, obtained thusly. A Gaussian fit is applied to the
histogram of offsets produced by accumulating APOLLO range errors reported
for ACS photon pairs as described above (see Figure~\ref{fig:gaussian}).
The range error is the Gaussian fit centroid, expressed in millimeters.
The estimated error of the centroid, $\mu$, is simply the Gaussian standard deviation $\sigma$ divided by
$\sqrt{N_{\mathrm{pairs}}}$.  As is apparent in Figure~\ref{fig:gaussian},
the fit is rather good. Reduced chi-squared ($\chi^2_\nu$) measures for all
fits used in the present analysis are statistically probable.  The spread
seen in ACS time differences in Figure~\ref{fig:gaussian}, at 8 TDC bins
(200~ps), indicates an intrinsic single-photon uncertainty down by
$\sqrt{2}$ from this, or 140~ps (21~mm).  This is consistent with other
metrics for APOLLO timing performance, but unfortunately requires large
numbers of photons in order to reach millimeter-levels. Before the ACS, we
could not be sure how much of this spread was attributable to the APOLLO
laser vs. APD detection  and other timing elements.  Since ACS photons are
generated in very short pulses, we now know that electronics influences
dominate, motiviating paths for future improvement.

Several key points should be made about the numbers in Table~\ref{tab:example-stats}.
First, the range errors are all tolerably small. This suggests
that APOLLO is not guilty of gross inaccuracy in its timing measurement.
Second, the numbers are not in statistical tension with
each other: relative to the weighted mean of range errors in
Table~\ref{tab:example-stats} of 0.355\,mm, the $\chi^{2}$
value would be larger for 19\% of random realizations using these
errors. Finally, when enough pairs are included in the comparison
(all channels, all shots, $\delta_{\mathrm{shot}}>0$), a single ACS+LLR
run is capable of sub-millimeter accuracy.

In a second test, the ACS accurately captures XL-DC clock drift---a roughly
1.5\,mm effect in this instance. In this game, we look at
smaller shot ranges and ask if we can discern the clock rate shift
in shots 7000--9000 compared to shots 1--7000, obvious in Figure~\ref{fig:lun_raw}.
We observe the first part of the run to accumulate 60 TDC bins of
phase offset in 7000 shots, or 1.5\,ns in 350 seconds, or $-4.3\times10^{-12}$
fractional frequency offset. Between shots 7000--9000, we see a reversal
of 16 TDC bins, translating to a rate of $4\times10^{-12}$. Across
the lunar round-trip time of 2.5\,s, we might therefore expect a time
measurement error of $-4.3\times10^{-12}\cdot2.5\cdot0.15\frac{\mathrm{mm}}{\mathrm{ps}}=-1.6$\,mm
and 1.5\,mm, respectively (a 3.1\,mm difference). Table~\ref{tab:shotrange}
presents the results. The first row indicates a negative offset for
the first period, different from the latter period (second row) by
about 3\,mm---perfectly in line with clock drift expectations two
sentences back. The two measures are about 4-$\sigma$ discrepant. 

\begin{table}[t]
  \caption{\label{tab:shotrange}Clock drift is easily seen/recovered ($\delta_{\mathrm{shot}}=5$,
all channels).}
  \begin{indented}
  \lineup
  \item[]\begin{tabular}{@{}lll}
    \br
    Shot range & $N_{\mathrm{pairs}}$ & Range error (mm)\\
    \mr
    \0\0\01--7000  & 6219 & $-0.31\pm0.38$  \\
    7000--9000     & 1767 & $\m2.69\pm0.73$ \\
    \br
  \end{tabular}
  \end{indented}
\end{table}

A third test reveals that effects other than clock drift are small. The
clock drift is due to a frequency offset, producing a range error
only when developed over a time interval: \eg, 2.5\,s. If we eliminate
this time interval by comparing fiducial and lunar gates that are
about 50 shots different (2.5\,s times 20 shots/s), they are effectively
simultaneous. In other words, fiducial shot 1000 will be compared
to contemporaneous lunar return numbered 950 ($\pm\delta_{\mathrm{shot}}$), exposing any range
error that persists from influences other than clock drift. Table~\ref{tab:Simult}
indicates statistically consistent ``other'' offsets with a weighted
mean of $1.44\pm0.34$\,mm. This error scale is in line with previous
APOLLO assessments based on comparing the performance of different
channels to each other, and may be due in part to EMI from the APOLLO laser firing. Comparing Table~\ref{tab:shotrange}
values to the weighted mean from Table~\ref{tab:Simult}, we verify
that the 1--7000 shot range has a clock-induced offset of $-1.75\pm0.51$\,mm
and the 7000--9000 shot range has an offset of $1.25\pm0.81$\,mm:
again consistent with expectations from the independent measurement
of clock frequency ($-1.6$ and 1.5\,mm). 

\begin{table}[t]
  \caption{\label{tab:Simult}Neutralizing clock offsets by evaluating essentially simultaneous
events, we see a consistent range error from sources other than the
clock ($\delta_{\mathrm{shot}}=5$, all channels).}
  \begin{indented}
  \lineup
  \item[]\begin{tabular}{@{}lll}
    \br
    Shot range    & $N_{\mathrm{pairs}}$ & Range error (mm)\\
    \mr
    \0\0\01--7000 & 5992 & $1.65\pm0.38$  \\
    7000--9000    & 1749 & $0.68\pm0.72$  \\
    \br
  \end{tabular}
  \end{indented}
\end{table}

\begin{table}[t]
  \caption{\label{tab:chan-off}Differential channel offsets (partial set; channel 15 is not used).}
  \begin{indented}
  \lineup
  \item[]\begin{tabular}{@{}lll}
    \br
    Channel & $N_{\mathrm{pairs}}$ & Range error (mm) \\
    \mr
    11 & 1225  & $-0.42  \pm 0.88$ \\
    12 & 1382  & $-1.30  \pm 0.76$ \\
    13 & 1403  & $\m2.11 \pm 0.81$ \\
    14 & \0998 & $\m6.60 \pm 0.92$ \\
    16 & 1174  & $\m3.75 \pm 0.81$ \\
    \br
  \end{tabular}
  \end{indented}
\end{table}

A final test shows that the ACS can nail down differential channel offsets in ten minutes,
compared to months in the past. In a single ACS-LLR overlay run, we get an
independent measure of two types of channel-specific timing offsets: large
offsets (1\,ns scale) common to both fiducial and lunar gates, due to
essentially static delays that are different for each channel; and
differential offset between gate types ($\lesssim 0.1$\,ns level). The
latter type of offset results in different lunar range measurements for
each channel, for instance. We have routinely corrected for these offsets in
the past, but adequate characterization required months of lunar data,
relying on particularly strong returns in order to have enough ``truth''
measurements from the Moon. Being able to measure the channel-dependent offsets
in a single run opens the door to studying long-term evolution and causes, which will be the subject of future work. 

Table~\ref{tab:chan-off} illustrates typical values of differential channel
offsets from this run---each row representing a different channel.  The
offsets are sometimes substantial, wholly inconsistent with each other, and
have decent precision (using $\delta_{\mathrm{shot}}=9$ and gate offset of
$-50$ to neutralize clock drift). Within an observing session, we tend to
see stable, self-consistent differential channel offsets, and little in the
way of worrisome structure across longer timescales. In short, ACS delivers
superior channel offset characterization, on \emph{much} shorter
timescales.

In summary, these example ACS data indicate that APOLLO inaccuracies
are substantially lower than $\gtrsim 15$\,mm model residuals;
we can characterize inaccuracies at the sub-millimeter
level; we no longer have to rely on lunar ranging for truth data,
but can produce it when we want; and we can commence an effort to correct our now-characterized few-millimeter-scale errors.  

\subsection{Range error statistics}

Section~\ref{sec:example-run} highlighted key analysis schemes available
using the ACS.  A future paper will describe additional tests and measures,
and elucidate the procedures by which ACS data are used in the APOLLO data
reduction to produce accurate LLR normal points.  For now, we apply the
techniques from Section~\ref{sec:example-run} to all APOLLO operations
since 2016 September 12 (through 2017 April 3) to understand overall performance
and stability.

\begin{figure}[t]
  \centering
  \includegraphics[width=12cm]{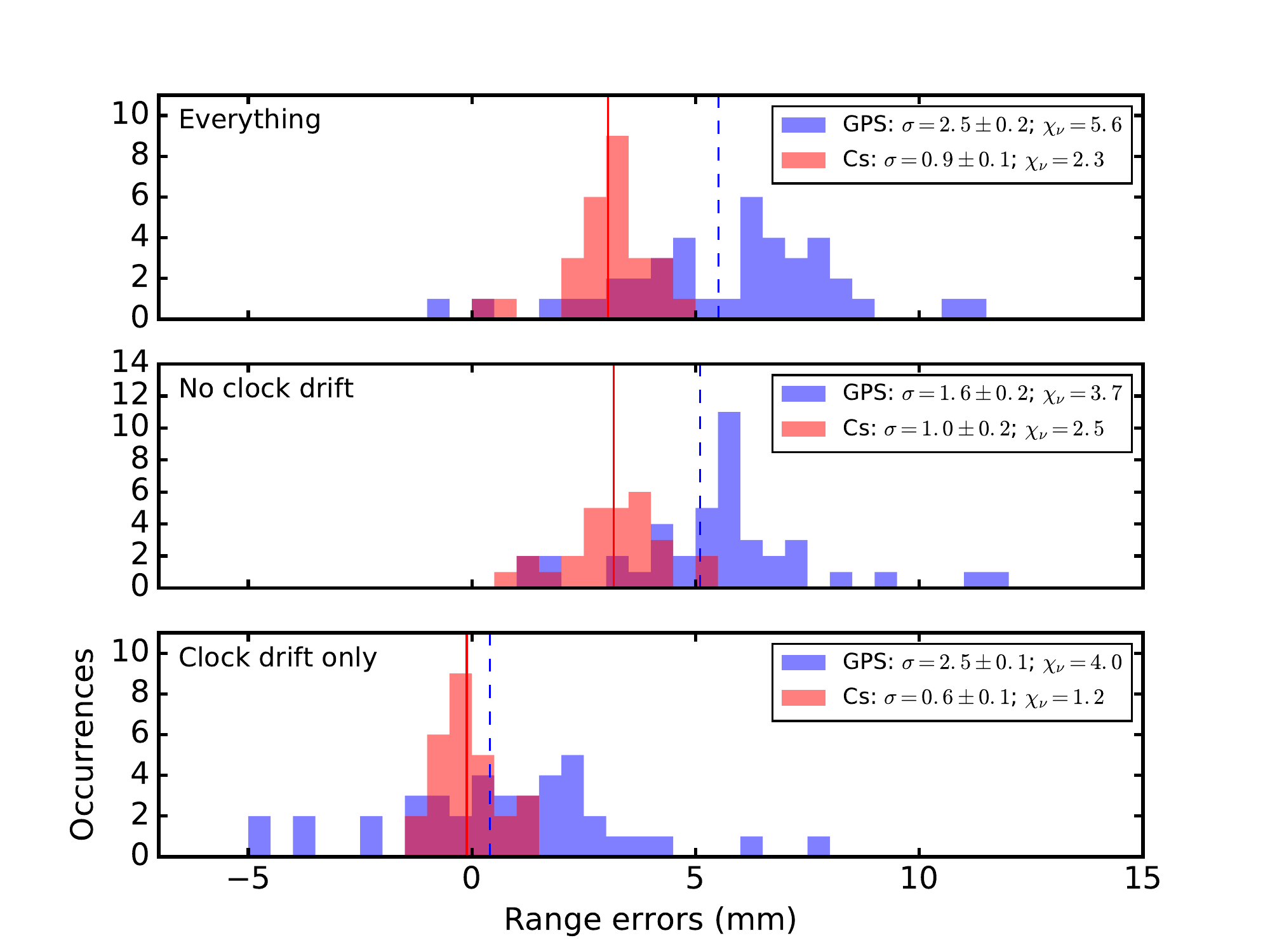}
	\caption{Channel-aggregated range offsets (0.5\,mm bins);
Blue/lighter: GPS-based; Red/darker: cesium-based data (darkest is overlap). The top
histogram compares ACS fiducial to lunar photons having the same shot
number and is an “everything” measure, containing systematic contributions
from timing electronics and clock effects; the middle histogram is the
non-clock measure (lunar shot offset by $-50$ shots comparing effectively
simultaneous events) and the last is the difference of the two previous
distributions, describing the clock-only contribution to range errors. All
cases use $\delta_{\mathrm{shot}} = 5$
Weighted means are represented by dashed (GPS) and solid (Cs)
vertical lines.
Weighted standard deviations and their computed errors are shown
for each. Reduced chi statistics describe how well the weighted mean fits
the distribution relative to the offset errors. A factor of four reduction in
clock-contributed range errors is apparent in the cesium-only spreads
relative to GPS data.\label{fig:aggrRangeOffs}}%
\end{figure}

Figure~\ref{fig:aggrRangeOffs} summarizes the results.  The top panel
corresponds to the measure presented in Table~\ref{tab:example-stats},
using $\delta_{\mathrm{shot}}=5$ and all channels.  The middle panel
corresponds to the ``simultaneous'' measure of
Table~\ref{tab:Simult}---eliminating the effect of clock drift (though not
clock jitter).  The top panel therefore represents all effects together,
and
the middle panel is everything but clock drift.  The final panel performs a
subtraction of the two to characterize the clock drift itself.

At the beginning of
2017, we switched the APOLLO frequency standard to the Cs clock (still
using the GPS clock as an absolute time reference).  We have therefore
split the analysis into two periods corresponding to reliance on the two
clocks.  Besides reducing phase noise (thus jitter), using the Cs clock as
the frequency standard guarantees no drift relative to the ACS laser
pulses, given the phase lock to the same clock. Thus the Cs-based data in
Figure~\ref{fig:aggrRangeOffs} appears to be much tighter.

Figure~\ref{fig:aggrRangeOffs} demonstrates that the GPS clock
contributes about 2.5~mm of spread in APOLLO range data.  Previous
estimates of clock-induced error were based on disciplining DAC steps
resulting in $1.2\times 10^{-11}$ fractional frequency shifts, translating
to 4.5~mm in one-way range. The stable DAC behavior suggested a roughly
uniform offset distribution bounded by a 4.5~mm range, so that the RMS
error would be a factor of $\sqrt{12}$ smaller, or roughly 1.3~mm.  The ACS
has revealed that the GPS clock drift is is about twice what we
anticipated---consistent with the clock comparison campaign alluded
to in Section~\ref{sec:clock_comparison} and treated extensively in a
separate paper \cite{APOLLO_clock2017}.  The reduced
$\chi^2$ measure for the Cs clock-drift-only distribution indicates no
spread beyond the uncertainties contributed by the ACS statistics
(corresponding to errors in Tables~\ref{tab:example-stats}, for instance).
Neither clock appears to contribute significant range bias (both zero-mean).
We do see a several-millimeter offset in the collective statistics (best
seen in middle panel of Figure~\ref{fig:aggrRangeOffs}).  We should not
take this result literally just yet, as a deeper analysis of ACS data
suggests that this is due to temporal structure likely stemming from the
APOLLO laser EMI.  A more thorough utilization of ACS data can
correct for this as well, to be detailed elsewhere.
We have also investigated the stability of channel offset measures, as
explored in Table~\ref{tab:chan-off} for the example run, finding the
behavior to be stable over month-long periods.

The ACS has provided a powerful check on APOLLO performance.  So far, the
indications are reassuring.  While it is true that historical APOLLO data
are more impacted by clock drift than we had appreciated, the level is not
severe (2.5~mm), and part of this is correctable \cite{APOLLO_clock2017}.  We also
see static offsets at the few-millimeter scale, but these may well
disappear upon a more complete analysis, as ongoing work suggests.

\section{Conclusions}

The ACS has transformed the way APOLLO collects LLR data.  While offering
reassurance that historical APOLLO data appear to be free of
large-scale systematic errors at the level of current model residuals
($\gtrsim 15$~mm), the
ACS \emph{has} exposed some systematic errors larger than previously
appreciated.  The GPS-disciplined clock contributed $\sim 3$\,mm of
random error---about double the previously-reasoned $\sim 1.5$\,mm scale.
Fortunately, the addition of a cesium clock (not to mention the ACS
capability as a whole) renders this point moot going forward.  A separate
paper demonstrates a method by which we can reduce this clock error for
historical APOLLO data \cite{APOLLO_clock2017}.
We also see a static
offset of several millimeters that was not previously measured, deserving further attention.  While such an offset
is habitually removed in the model via a constant range bias, it is preferable to
understand and remove any such influences, especially if the offset is found to evolve
over time.  While not stressed as much here, the ACS also provides a powerful
and fast diagnostic tool to help characterize and eliminate
systematic error influences.  

A likely outcome of this effort is increased pressure on model improvement,
having largely settled the question of APOLLO data quality.  In the past,
APOLLO data were not as accurate as the estimated uncertainties indicated,
although not by enough to explain model residuals.  Sharpening up APOLLO
data may more clearly expose systematic signatures in the model, suggesting
routes to improvement.

In the end, we look forward to realizing the potential gains of
millimeter-accurate LLR measurements.  Order-of-magnitude gains in our
understanding of fundamental gravity are a compelling reward.

\subsection*{Acknowledgments}

We thank Ed Leon for performing periodic measurements of clock phase
at the observatory. George Kassabian provided a design for producing
well-formed RF pulses for the EOM with adjustable amplitude.  Jim
MacArthur and James Phillips provided assistance with other aspects of
the LSB design. This work is based on access to and observations with
the Apache Point Observatory 3.5-meter telescope, which is owned and
operated by the Astrophysical Research Consortium.  This work was
jointly funded by the National Science Foundation (PHY-1404491) and
the National Aeronautics and Space Administration (NNX-15AC51G).

\section*{References}
\bibliographystyle{iopart-num}
\bibliography{apollo}

\end{document}